# Algorithmic Administration and the EU AI Act: Legal Principles for Public Sector Use of AI

*Ioannis Kastanas\*1, Georgios Pavlidis (corresponding author)\*2*

*\*Jean Monnet Center of Excellence AI-2-TRACE-CRIME, Neapolis University Pafos*

The increasing use of artificial intelligence (AI) by public authorities introduces both opportunities for innovation and significant challenges for the administrative rule of law. This article examines how the EU AI Act interacts with the fundamental principles of administrative law, with a particular focus on administrative discretion, the duty to state reasons, and the principle of proportionality. The article analyses the regulatory obligations imposed by the AI Act on public sector deployers of high-risk systems, especially in sensitive domains such as social benefits, migration, education, and law enforcement. It also explores whether the AI Act adequately ensures accountability, transparency, and reviewability in automated public decision-making. The article further considers how the AI Act's risk-based approach aligns (or fails to align) with the principle of proportionality and it proposes safeguards and interpretative strategies to ensure the ethical and lawful deployment of AI in the public sector.

**Acknowledgement:** This study was prepared in the context of the Jean Monnet Center of Excellence AI-2-TRACE-CRIME and was funded by the European Union. Views and opinions expressed are however those of the authors only and do not necessarily reflect those of the European Union or the European Education and Culture Executive Agency (EACEA). Neither the European Union nor EACEA can be held responsible for them.

---

[1] Lecturer of Public Law, Team Member of the Jean Monnet Center of Excellence AI-2-TRACE-CRIME, Neapolis University Pafos, Cyprus, Email: i.kastanas.9@nup.ac.cy

[2] Associate Professor of International and EU Economic Law, UNESCO Chair & Jean Monnet Chair, School of Law, Neapolis University Pafos, Cyprus, Email: g.pavlidis@nup.ac.cy

## 1. Introduction

The deployment of artificial intelligence (AI) systems in public administration has the potential to transform the exercise of administrative authority and the delivery of public services. From welfare benefits and tax audits to immigration control and predictive policing, AI tools are being adopted to enhance administrative efficiency and to improve resource allocation in the public sector (Agarwal, 2018). The use of AI promises to automate decision-making processes in public administration, manage large datasets in the public sector, and deliver services at scale more effectively. Yet this algorithmic turn in the public sector has also raised ethical and legal, even constitutional, concerns. High-profile failures—such as the Dutch childcare benefits scandal (Toeslagenaffaire), where risk-scoring algorithms falsely labelled thousands of families as fraudsters, or the use of opaque decision-support tools in procedures of asylum— have demonstrated how public administrations, their personnel, and their technology can reproduce or exacerbate existing biases. Such practices may erode public trust in institutions and cause significant harm to individuals, particularly those in vulnerable situations.

The European Union (EU) has sought to respond to these concerns through the AI Act,[3] adopted in 2024. The Act which establishes a risk-based regulatory framework for the development, placing on the market, and use of AI systems. It explicitly includes within its scope high-risk systems used by public authorities, such as those affecting access to education, employment, welfare, law enforcement, and migration. It imposes specific obligations not only on providers of AI systems, but also on deployers, which can be public bodies. Among these obligations, the AI Act requires providers and deployers of AI to ensure meaningful human oversight, maintain detailed documentation, and uphold fundamental rights.

This article investigates the deployment of AI in public administration through the lens of administrative law. It explores how the AI Act regulates the use of AI systems by public authorities, the extent to which algorithmic decision-making challenges core principles of administrative law and whether the safeguards introduced by the AI Act offer sufficient protection for individuals subject to automated decisions.

---

[3] Regulation (EU) 2024/1689 of the European Parliament and of the Council of 13 June 2024 laying down harmonised rules on artificial intelligence (Artificial Intelligence Act), OJ L, 2024/1689, 12.7.2024,

## 2. Automated Decision-Making in Public Administration: An Administrative Law Perspective

Public administrations worldwide have gradually adopted technological tools, establishing regulatory frameworks for automated administrative procedures that, where possible, replace traditional processes, driven by the pursuit of the public interest and the need for more efficient public services (Lazaratos, 1990). The integration of AI systems into public administration engages two sets of legal requirements: those governing the deployment of AI in administrative decision-making and those concerning the development and training of such systems. Deployment-related obligations stem from the foundational tenets of administrative law—legality, transparency, proportionality, participation, and accountability—which are rooted in both national traditions and EU law (Van Noordt and Tangi, 2023). While these legal principles place limits on AI use in particularly sensitive areas, the current legal framework remains permissive towards the deployment of AI by public authorities (Krönke, 2025). For their part, the development-oriented requirements, especially those enshrined in the AI Act, are more prescriptive. The AI Act introduces harmonised rules for AI systems, including specific safeguards for high-risk AI applications and general-purpose AI models, and prohibits certain harmful practices. Its objective is to promote trustworthy and transparent AI, which in the context of public administration must also respect some key principles of administrative law.

First, the principle of legality requires that administrative decisions have a basis in law and conform to applicable legal standards (Meneceur, 2023). This includes both procedural requirements and substantive limits on discretion. Algorithmic systems, however, often operate in ways that obscure the legal source or justification of decisions. This is particularly true where the model logic is inaccessible or relies on opaque data processing steps. This raises the question as to whether decisions taken (or informed) by AI systems can be considered lawful in the traditional sense of administrative law. For example, the deployment of predictive policing tools without a clear legal mandate and explicit statutory basis raises concerns that administrative actions based on the resulting algorithmic risk assessments may lack the lawful authority required under the principle of legality.

Alongside legality, transparency remains another fundamental principle of administrative law—one that digital transformation in governance must continue to respect and preserve (Sharmin and Chowdhury, 2025). Transparency encompasses both access to the relevant rules and criteria, as well as transparency about how a specific decision was reached in a specific situation. In the context of AI, transparency is frequently hindered by tech-nical opacity—commonly referred to as the "black box" problem (Pavlidis, 2024)—as well as by trade secret protections and the inherent complexity of machine learning models. The lack of transparency not only affects the right to be heard and to challenge decisions but also undermines public trust. A notable example is the French government's use of the Parcoursup platform for university admissions, where the opacity of the algorithmic selection criteria drew criticism for lacking sufficient transparency and accountability (Cluzel-Métayer, 2020). This has raised concerns that administrators and public officials may increasingly defer to algorithmic outputs without adequate justification, effectively reducing their explanations to little more than a reference to the system's decision, thereby undermining the principle of reasoned decision-making. To reduce opacity in the use of algorithms, the French Administrative Code requires informing the administered individuals about the degree and manner of algorithmic processing's contribution to decision-making, the data being processed and its sources, the parameters of this processing, and finally its functions. It should be noted that this obligation of detailed information exists even in cases where AI merely contributes to decision-making and does not replace the deciding body or person (Geburczyk, 2021; Edwards and Veale, 2017).

Moreover, the principle of proportionality, rooted in both EU and national public law, requires that administrative measures be appropriate and necessary to achieve legitimate aims and not excessively burdensome (Warthon, 2024). Algorithmic systems, particularly those designed to optimise for efficiency or risk reduction, may overlook the broader social impacts of decisions. Excessive reliance on algorithmic scoring or classification tools may lead to disproportionate outcomes in individual cases. A real-world example is the use of automated fraud detection systems in welfare administration—such as the Dutch Toeslagenaffaire—where minor discrepancies or risk indicators triggered severe sanctions, including the withdrawal of benefits, without adequate individual assessment. Similarly, in a hypothetical scenario, an AI-driven housing allocation system operated by a public authority or local government that systematically deprioritises applicants with irregular employment histories—without accounting for individual circumstances or offering procedural safeguards—could lead to unjust and excessive exclusion from access to essential public services, thereby violating the principle of proportionality.

Alongside the principles mentioned above, participation refers to the rights of individuals to be involved in administrative processes that affect them, whether through consultation, objection, or submission of evidence (Wong and others 2025). AI tools may reduce the opportunity for meaningful participation, especially where automated processes are presented as final or non-negotiable. This is particularly problematic in welfare, immigration, or social service contexts, where individuals are often in a structurally weaker position. An illustrative example is the use of automated systems in visa processing, where individuals often face significant difficulties in understanding or contesting negative decisions due to the absence of meaningful explanations or accessible appeal mechanisms. In some cases, such as the Australian visa cancellation regime, what emerges may be a form of "surveillance fantasy"—a perception that deportation decisions are fully automated, whereas in reality, discretion remains with human decision-makers. Nevertheless, even where full automation is not in place, these systems continue to raise concerns about efficiency, accuracy, and fairness (Weber and Gerard, 2024). In another example, if an AI system used to allocate disability benefits were to automatically reject applications based on rigid eligibility thresholds without offering applicants the chance to provide contextual medical or social information, it would significantly curtail the right to be heard and diminish procedural fairness.

Finally, accountability entails that public authorities can be held responsible for their actions, including through administrative review, judicial control, and political scrutiny (Bracci, 2023; Yuan and Chen, 2025). Yet algorithmic systems introduce a certain diffusion of responsibility. Indeed, decisions may result from a combination of human and automated processes, making it difficult to identify who is accountable, especially when the system's provider, not the deployer, controls its design and logic. A real-life example is the UK's A-level grading controversy in 2020, where a standardisation algorithm downgraded thousands of students' scores, leading to widespread public backlash and institutional confusion over who was ultimately responsible—the government, the exam regulator, or the developers of the algorithm (Hughes, 2020). In similar cases, where an AI system erroneously takes a decision, affected individuals may face significant difficulties in determining where accountability lies—whether with the public authority that deployed the tool, the software vendor responsible for its design, or the entity that supplied or curated the training data.

The jurisprudence of the European Court of Human Rights (ECtHR) has begun to engage with the challenges posed by automated decision-making, albeit in a limited and indirect manner. In its landmark Grand Chamber case Roman Zakharov v. Russia (2015),[4] the ECtHR ruled that Russia's legislation enabling broad surveillance powers—facilitated by automated interception tools—violated Article 8 ECHR (private life). The Court criticised the absence of clear legal constraints, judicial oversight, and effective remedies. This highlights how automated systems can lead to arbitrary or secretive public authority actions without accountability. In López Ribalda v. Spain (2019),[5] the ECtHR further underscored the importance of transparency and necessity in the use of surveillance technologies in the workplace. In Toplak and Mrak v. Slovenia (2021),[6] although revolving around accessible voting machines in public elections, the ECtHR emphasised state obligations to ensure meaningful participation and accountability when deploying technology in democratic processes. This case reinforces that public automation tools must be inclusive, reviewed, and responsible.

---

[4] ECtHR, Roman Zakharov v. Russia (Application no. 47143/06), judgment of 4 December 2015.

Thus far, the Court of Justice of the European Union (CJEU) has not addressed the legal implications of automated decision-making in a direct and comprehensive way. In La Quadrature du Net,[7] for instance, the CJEU held that national provisions allowing for the general and indiscriminate retention of traffic and location data for national security purposes are, in principle, precluded under EU law, except in narrowly defined circumstances. The Court thereby underscored the importance of proportionality and effective oversight in the use of algorithmic surveillance and data processing by public authorities, particularly in the context of law enforcement. While such cases do not directly concern administrative decision-making in the traditional sense, they reflect broader concerns about due process and the protection of fundamental rights in the digital age. With the entry into force of the AI Act, it is anticipated that the CJEU will, in time, be called upon to interpret its provisions—especially those concerning high-risk and general-purpose AI systems—thereby contributing to the development of a more coherent and detailed case law on the legality, proportionality, and accountability of AI use in public administration.

Moreover, national courts across Europe have started scrutinising the use of algorithms in public decision-making. A notable example is the 2020 judgment of the District Court of The Hague in the Netherlands,[8] which struck down the SyRI (System Risk Indication) system—an algorithmic profiling tool used to detect social fraud—on the grounds of insufficient transparency and a lack of adequate safeguards (Appelman, Fathaigh, and van Hoboken, 2021). Similarly, in the Aerius case, the Dutch Council of State held that public authorities have a duty to actively and promptly disclose the fundamental choices underlying the deployment of automated decision-making systems.[9] Such developments demonstrate that the deployment of AI systems will pose serious challenges to the application of long-established administrative law principles. While AI may deliver substantial gains in efficiency and resource allocation, these benefits must not come at the expense of core values such as legality, transparency, proportionality, participation, and accountability. As the following section will explore, the AI Act seeks to offer regulatory responses to these concerns—though important questions remain as to whether it goes far enough in addressing the complexities of algorithmic governance in public administration.

---

[5] ECtHR, López Ribalda v. Spain (Applications nos. 1874/13 and 8567/13), judgment of 17 October 2019.
[6] ECtHR, Toplak and Mrak v. Slovenia (Applications nos. 34591/19 and 42545/19), judgment of 26 October 2021.
[7] Judgment of the Court (Grand Chamber) of 6 October 2020, Joined Cases C-511/18, C-512/18 and C-520/18, ECLI:EU:C:2020:791
[8] The Hague District Court, Judgment of 5 February 2020, Case number: C/09/550982 / HA ZA 18-388, available at: https://uitspraken.rechtspraak.nl/details?id=ECLI:NL:RBDHA:2020:1878

## 3. The AI Act and Public Sector Use of High-Risk AI Systems

The EU Artificial Intelligence Act (AI Act)[10] introduces a risk-based regulatory framework that classifies AI systems into four categories: unacceptable risk, high risk, limited risk, and minimal risk. AI systems are prohibited when they present unacceptable risk, while high-risk AI systems are subject to the most stringent obligations. A significant number of high-risk systems are likely to be used by or on behalf of public authorities, especially in areas where administrative decisions have significant effects on individuals' rights and legal interests.

### 3.1 Overview of Obligations on Deployers

Under the AI Act, deployers of high-risk AI systems are required to implement technical and organisational measures to ensure usage strictly adheres to the provider's instructions and maintain competent human oversight throughout operation; they are responsible for ensuring input data is relevant and representative, continuously monitor system performance, and promptly report any identified risks or serious incidents to the provider, importer, distributor and relevant authorities—suspending use when necessary.

Importantly, prior to deploying a high-risk AI system, Article 27 imposes an obligation on deployers that are bodies governed by public law, or are private entities providing public services, and deployers of such systems to perform an assessment of the impact on fundamental rights that the use of such system may produce. This requirement covers deployment of AI by public authorities, including courts and law enforcement bodies and reflects an acknowledgment that public sector use of AI affects rights in a particularly sensitive way (Mantelero, 2024). The assessment must address the impact of the AI system on privacy, non-discrimination, freedom of expression, and other fundamental rights protected under the Charter of Fundamental Rights of the EU. This aligns with traditional administrative law requirements of legality and accountability but introduces a novel *ex ante* assessment that is not common in many Member States' administrative practices. It also seeks to close the accountability gap between private developers and public users of AI tools.

---

[9] See ABRvS (Judicial Division of the Council of State) 17 May 2017, ECLI:NL:RVS:2017:1259 (Aerius I); ABRvS 18 July 2018, ECLI:NL:RVS:2018:2454 (Aerius II).
[10] Regulation (EU) 2024/1689 of the European Parliament and of the Council of 13 June 2024 laying down harmonised rules on artificial intelligence (Artificial Intelligence Act), OJ L, 2024/1689, 12.7.2024,

### 3.2 Categories of High-Risk Systems Relevant to the Public Sector

Article 6(2) of the AI Act and its Annex III define a broad set of high-risk AI systems, many of which directly concern public administration activities.[11] These include systems used in biometrics, critical infrastructure, education, employment, access to essential public services, law enforcement, migration and border control, and the administration of justice and democratic processes. All these categories reflect areas where public authorities, institutions, or agencies may deploy AI in ways that significantly affect individuals' rights and interests. Each of these domains involves the exercise of public power with potential consequences for rights and legal status. Importantly, the risk classification under the AI Act does not depend solely on the technical design of the system but on the context and purpose of use (De Cooman, 2022)—which means that identical systems used in the private sector may not be high-risk, whereas their public sector deployment is. This distinction reflects the constitutional dimension of public authority: when the State acts, it must do so under a higher standard of justification, given its monopoly on coercive power.

A more nuanced view is offered by Article 6(3) of the AI Act, which stipulates that the classification of a system as high-risk is not purely formalistic but must also reflect the actual use and the potential consequences of deployment. In other words, a system listed in Annex III does not automatically become high-risk in every conceivable application: it must be used for the intended purpose in a context that engages fundamental rights or safety concerns. This provision demonstrates that the classification is based not only on the system's design but also on its real-world application and the risks it entails. Interestingly, however, the Act does not recognise the inverse situation. Where an AI system is not formally classified as high-risk but is nevertheless used in a sensitive context with potentially severe implications for rights, the Act does not automatically extend the high-risk obligations. This asymmetry seems to leave a potential regulatory gap: the law attempts to prevent over-classification of low-stakes applications, but it seems to under-regulate situations where seemingly low-risk tools have profound impacts. This situation confirms the importance of the principle of proportionality and the need for contextual evaluation in administrative law as a complementary interpretative tool.

---

[11] See also Recitals 48, 52 and 54-63 of the AI Act.

### 3.3 The Notion of "Meaningful Human Oversight" and Its Legal Implications

A cornerstone of the AI Act's high-risk regime is the requirement of meaningful human oversight, as outlined in Article 14 of the AI Act. This concept, although technologically and ethically complex, is essential for preserving legality, discretion, and accountability in public administration. Human oversight must be more than formal or symbolic. The human overseeing the system must understand the AI system's functioning and limitations, be in a position to challenge or override, reverse or stop outcomes where necessary, and have access to adequate documentation and support to intervene appropriately.

From a legal perspective, meaningful human oversight acts as a procedural safeguard—it creates a point of contact between automated processing and administrative responsibility. However, challenges persist, particularly in the public administration where oversight mechanisms can be under-resourced. In such cases, oversight may devolve into mere formality, especially where officials defer to system outputs due to their perceived authority or complexity. This is a well-known phenomenon, known as automation bias (Alon-Barkat and Busuioc, 2023). Moreover, in situations involving large-scale or batch processing (e.g. benefit allocation), it may be practically difficult (or impossible) for human officials to adequately review every case, raising doubts about the effectiveness of the safeguard in practice. Future guidance from the European Commission, the AI Office, and potentially national courts would be valuable in promoting uniform standards of administrative fairness in this context.

## 4. Administrative Discretion and Algorithmic Constraint

The use of AI systems in public administration has implications for the exercise of administrative discretion—the space within which public authorities can interpret and apply legal norms and tailor decisions to individual circumstances. While discretion is not unfettered, it plays a central role in modern governance. Discretion enables public bodies to adapt general rules to complex and context-sensitive situations (Shapiro, 1983). The integration of AI into decision-making processes, however, introduces new forms of constraint that may narrow the exercise of such discretion, with important consequences for the rule of law and individual rights (Barth and Arnold, 1999).

### 4.1 How AI May Narrow or Obscure Administrative Discretion

AI systems, particularly those using machine learning, typically rely on historical data to identify patterns, predict outcomes, or classify cases. In doing so, they tend to standardise decision-making, in other words to treat similar inputs in a uniform manner. While this can promote consistency, it may also pre-structure decisions, thereby reducing the space available for case-by-case assessment. Where public officials are expected (or encouraged) to follow the output of an AI system, their ability to exercise independent judgment may be limited. Moreover, the design choices embedded in AI models—such as which variables are considered relevant, how they are weighted, and how uncertainty is treated—often remain invisible to public officials that are the end-users. This "design discretion" effectively displaces administrative discretion from public officials to system developers, who may operate under very different incentives and lack democratic legitimacy (Beckman, Rosenberg, and Jebari, 2024). As a result, the locus of discretion shifts. In some cases, AI systems may also obscure discretion by presenting outcomes as deterministic or inevitable, with an appearance of objectivity. For instance, risk scores generated in the context of fraud detection or welfare eligibility may be perceived as objective assessments, even if they rest on probabilistic modelling. This can lead to automation bias (Ruschemeier and Hondrich, 2024), whereby human officials defer to the system even when they retain formal decision-making authority.

### 4.2 The Duty to Give Reasons and the Explainability of Algorithmic Decisions

One of the core guarantees of administrative fairness under EU law and the laws of Member States is the duty to give reasons for administrative de-cisions. This requirement serves several purposes: it enhances transparency, facilitates judicial review, and enables affected individuals to understand and contest decisions. When AI systems play a decisive role in administrative processes, the extent to which meaningful reasons can be given becomes limited. In particular, explainability—the ability to provide a comprehensible account of how a decision was reached—faces challenges in the context of complex or opaque models (Angelov and others, 2021).

While the AI Act requires that high-risk systems be designed to enable interpretability and auditability (Article 13), it stops short of requiring full transparency about the model's inner workings. Article 86 introduces a general right to explanation in relation to the use of AI; however, its contours remain under-defined and seem to lack specificity and operational clarity. (Nnawuchi and Carlisle, 2024). This is especially problematic in cases involving black-box algorithms, where the logic linking input data to outputs may not be readily intelligible even to the developers themselves. This difficulty reflects a deeper ambiguity identified in the doctrine (Sarra, 2019): what is presented as 'knowledge' in algorithmic outputs is never neutral from a normative point of view, but it rests on interpretive and design choices that have legal significance. In administrative contexts, this means that opacity in AI systems is not merely a technical concern but one that directly affects the intelligibility of decisions and the ability of individuals to exercise their right to good administration.

It can be argued that the duty to give reasons must evolve to include model-level explanations (how the system works in general) and instance-level explanations (why a particular output was generated in a specific case). Yet the technical feasibility and normative adequacy of such explanations remain open questions. In the absence of sufficient explainability, the right to good administration under Article 41 of the EU Charter may be undermined (Bousta, 2013).

## 4.3 Impacts on Fair Decision-Making, Particularly for Vulnerable Populations

In the Dutch childcare benefits scandal, briefly mentioned earlier, approximately 26,000 parents were wrongly accused of fraudulently claiming benefits between 2005 and 2019. As a result, they were ordered to repay the full amount of the allowances received, which in many cases reached tens of thousands of euros and plunged families into severe financial distress. Investigations revealed that the procedures followed by the Tax and Customs Administration were discriminatory, disproportionately targeting parents with migrant backgrounds and exhibiting systemic institutional bias. The scandal's gravity led to the resignation of the third Rutte cabinet on 15 January 2021, just two months ahead of the scheduled general election. A subsequent parliamentary inquiry concluded that the affair constituted a grave violation of fundamental rule of law principles.[12] This case illustrates that the delegation of discretion to algorithmic systems can also have disproportionate impacts on vulnerable groups, such as welfare recipients, asylum seekers, or individuals with limited digital literacy (Hadwick and Lan, 2021). These populations are more likely to be subject to automated processing and less likely to be in a position to challenge or understand such decisions. Algorithmic systems may encode and perpetuate structural biases present in training data or institutional practices (Mishra and others, 2024). For example, if past data reflect discriminatory enforcement patterns, predictive models trained on that data may replicate those patterns. This can happen even in the absence of explicit bias. Where discretion is constrained by such systems, there may be fewer opportunities to detect and correct injustices. This is particularly problematic in environments with limited legal assistance or judicial oversight. Moreover, vulnerable individuals may lack the procedural knowledge or resources needed to navigate complex appeals processes. Ultimately, the use of AI in public administration might transform discretion from a site of responsive judgment into one of pre-programmed automation, unless safeguards are put in place. Such safeguards must ensure that discretion remains a site of human, accountable, and reasoned decision-making.

## 5. Risk-Based Regulation and Proportionality in Administrative Enforcement

The EU AI Act adopts a risk-based approach to regulation, whereby legal obligations increase in proportion to the potential harm that an AI system may cause to health, safety, or fundamental rights. While this model is coherent with broader EU regulatory strategies—such as those in financial services, anti-money laundering or data protection—it raises important questions when applied to public administration, where principles of proportionality, equality, and fairness are already embedded in law. In this context, it is worth examining the interaction between risk-based regulation and administrative proportionality.

---

[12] Report of the Childcare Allowance Parliamentary Inquiry Committee entitled "Unprecedented injustice" (Ongekend onrecht) 17 December 2020, Parliamentary document 35 510, no. 3.

## 5.1 Proportionality in Administrative Law vs. Risk-Based Regulation under the AI Act

Proportionality is a general principle requiring that any measure taken by a public authority be suitable to achieve a legitimate aim, necessary in the sense of being the least restrictive option, and balanced in its effects (Thomas, 2000). By contrast, the AI Act's use of proportionality is system-centred, focusing on the risk that a particular AI system poses, rather than the proportionality of its deployment in a specific administrative context. This may lead to a misalignment between general risk levels and the contextual assessment required in public law. An AI system deemed "high-risk" under the Act may be used in a relatively low-stakes administrative procedure, raising questions about whether the associated regulatory obligations are excessive.

For instance, under Annex III of the AI Act, AI systems used in the context of access to public benefits and services (Annex III, point 5(a)) are classified as high-risk. This would include an AI tool used by a municipality to assist in processing applications for public library cards or minor housing subsidies. While technically falling within the high-risk category, such applications may involve low-stakes, routine decisions with limited impact on individuals' rights. Conversely, an AI system not explicitly listed in Annex III could be used in a sensitive administrative context, such as case prioritisation in child protective services, where it may significantly affect fundamental rights, yet escape the full scope of high-risk regulation if not captured by the enumerated use cases. Thus, the AI Act's thresholds may not always capture the real-life distribution of risk and harm. A purely technological risk assessment may miss these dimensions.

## 5.2 Are Obligations Proportionate to the Risks in the Public Sector?

The AI Act imposes a broad set of obligations on high-risk systems, including documentation, data governance, human oversight, and post-market monitoring. While these requirements aim to mitigate the specific risks associated with AI, they also carry administrative and financial costs—particularly for smaller public bodies, municipalities, or under-resourced agencies (Hoffmann and Nurski, 2021). The burden of compliance may be disproportionate in certain settings, especially where the AI system is used for non-discretionary or routine administrative tasks. For example, an AI system supporting appointment scheduling in a hospital or triaging routine tax returns may technically fall within a high-risk category but pose minimal real-world harm. In such cases, the proportionality of obligations should be assessed not only in relation to the system's abstract risk profile, but also in terms of actual use and context. Conversely, some deployments may be under-regulated. If a system escapes the high-risk classification because its primary function falls outside Annex III, but it is used in a way that affects fundamental rights (e.g. through indirect profiling or scoring), the protections of the AI Act may be insufficient. This creates a regulatory blind spot. Therefore, the principle of proportionality should inform not only the design of the AI Act's obligations, but also their application and interpretation by supervisory authorities. Sector-specific guidance could help ensure that the framework supports both innovation and rights protection in public administration.

### 5.3 The Role of Conformity Assessments and Post-Market Monitoring

Under the AI Act, conformity assessments are an important mechanism to ensure that high-risk AI systems comply with legal requirements before being placed on the market or put into service. For public authorities that develop or significantly modify AI systems in-house, this involves conducting an internal conformity assessment, including the preparation of technical documentation and a quality management system. While such procedures aim to prevent harm *ex ante*, they may be difficult to implement in public administrations that lack technical expertise or legal support. Moreover, conformity assessments may be treated as check-the-box exercises, focusing on formal compliance rather than actual impacts. This might reproduce the weaknesses of impact assessment regimes in other areas, such as data protection, where Data Protection Impact Assessments (DPIAs) are often underused (Demetzou, 2019).

The AI Act also introduces post-market monitoring obligations, requiring deployers to report serious incidents, monitor system performance, and keep logs for auditing purposes. These obligations are crucial for detecting unexpected harms; nevertheless, their effectiveness depends on the institutional capacity of public bodies and the existence of meaningful enforcement by national supervisory authorities and the European AI Office. Finally, proportionality should guide enforcement actions: sanctions or corrective measures under the AI Act must be tailored to the severity of the breach, the nature of the public body, and the impact on affected individuals. A mechanical application of enforcement powers could deter the legitimate use of AI in the public sector, especially where public interest justifications exist.

### 6. Safeguards and Legal Remedies

As AI systems are increasingly used to support or automate public sector decision-making, the question of how individuals can contest, review, or influence these processes becomes central. Legal safeguards and remedies must ensure that algorithmic decision-making remains compatible with the rule of law, fundamental rights, and principles of good administration. This section examines three core areas: the right to challenge algorithmic decisions, the need for institutional review mechanisms, and the role of transparency, complaint procedures, and participatory safeguards in supporting accountable algorithmic governance.

**6.1 The Right to Challenge Algorithmic Decisions**

The right to an effective remedy is enshrined in Article 47 of the Charter of Fundamental Rights of the European Union and in the legal orders of all Member States. Where administrative decisions are based on (or significantly influenced) by AI systems, individuals must retain the ability to challenge them, including on grounds related to the functioning, fairness, or legality of the algorithm. However, exercising this right in practice can be difficult. Individuals may not know that an algorithmic system was used in the first place, particularly in the absence of explicit notification. Even where such notification exists, the complexity of the system and the opacity of the logic behind the outcome may hinder effective legal challenge. As discussed earlier, explainability is under-defined and limited, especially in systems using complex or proprietary models. For its part, Article 50 of the AI Act addresses this issue only partly. It requires providers to ensure that 'AI systems intended to interact directly with natural persons are designed and developed in such a way that the natural persons concerned are informed that they are interacting with an AI system, unless this is obvious from the point of view of a natural person who is reasonably well-informed, observant and circumspect, taking into account the circumstances and the context of use' While this is a step forward, it falls short of providing a right to receive a meaningful explanation or a detailed account of the reasoning behind the decision. Moreover, the AI Act does not establish specific rights of appeal or redress beyond existing national and EU law frameworks. As such, the enforcement of rights will continue to depend largely on national administrative and judicial procedures, which vary in effectiveness.

Beyond the transparency duties of Article 50, the AI Act itself acknowledges a broader principle of contestability. Article 86 introduces a general right to explanation in relation to the use of AI. This demonstrates the EU's recognition that effective remedies depend on intelligibility of outcomes. The contours of Article 86 remain under-defined and seem to lack specificity and operational clarity (Nnawuchi and Carlisle, 2024). However, this right establishes a bridge between algorithmic accountability and the right to good ad-ministration under Article 41 of the Charter. Moreover, in cases where high-risk AI systems process personal data, the AI Act seems to operate in tandem with the GDPR. For its part, Article 22 GDPR prohibits decisions based solely on automated processing that significantly affect individuals, unless suitable safeguards are in place, including "the right to obtain human intervention, to express their point of view, and to contest the decision." This can be interpreted as linking the right to challenge to the right to a meaningful explanation of the underlying reasoning. The interplay between the AI Act and the GDPR creates a certain 'dual layer' of protection, which needs to be balanced by national authorities and courts (Sarra, 2025). For public administration, this means that automated decisions cannot be insulated from review, but they must remain open to contestation both procedurally and substantively.

### 6.2 Institutional Review Mechanisms

Effective oversight of algorithmic systems in the public sector requires not only individual remedies but also institutional review mechanisms capable of auditing systems and enforcing compliance. The AI Act foresees a multi-level governance structure. First, this structure involves national supervisory authorities, designated by Member States, responsible for market surveillance, enforcement, and guidance at national level. Second, it includes the European AI Office, established within the Commission, tasked with coordination for certain cross-border or high-impact AI systems. Finally, it involves sectoral regulators, which may be competent for specific domains such as data protection. In addition, public authorities using high-risk AI systems must conduct fundamental rights impact assessments, which can serve as an internal review tool, provided they are implemented effectively and subject to public or independent scrutiny. At the judicial level, courts remain the ultimate guarantors of legality, even in cases involving algorithmic decisions, though public interest litigation in this area is still relatively limited. To enhance oversight, mechanisms such as algorithmic audits (Le Merrer, Pons, and Trédan, 2024), ombudsman-led reviews, and possibly independent expert panels could be explored. The development of such hybrid oversight models could help build institutional capacity in this domain.

### 6.3 Transparency Registers, Complaint Procedures, and Participatory Safeguards

Transparency is a foundational condition for the exercise of rights, as well as for the functioning of democratic oversight. To that end, the AI Act introduces transparency-enhancing tools that could support greater accountability in public sector use. In addition to transparency obligations imposed on deployers, a public EU database of high-risk AI systems (Articles 49 and 71 of the AI Act) is to be established, listing systems placed on the market, including those used in the public sector. This can help civil society, journalists, and researchers monitor the use of AI in government. Under Article 27 and 8, Member States must establish accessible complaint mechanisms, allowing individuals or organisations to report non-compliance or harms resulting from AI deployment.

More broadly, there is a growing recognition of the need for participatory safeguards in the design, procurement, and deployment of AI in public administration. This includes public consultations prior to the adoption of AI systems, inclusion of affected communities in impact assessments, support for digital rights NGOs and independent auditors, and transparency obligations tailored to the informational needs of laypersons, not just experts. While such mechanisms are not (yet) fully mandated under the AI Act, they represent promising avenues for aligning AI governance with democratic legitimacy and public accountability.

## 7. Conclusion and Recommendations

AI systems have the potential to be deployed by public authorities to enhance efficiency, standardise decision-making, and manage complex bureaucratic tasks. At the same time, such systems risk undermining foundational principles of European administrative law—most notably legality, transparency, discretion, participation, and accountability—particularly when used in high-risk areas such as welfare, law enforcement, or migration control. The AI Act introduces several safeguards, including *ex ante* obligations such as fundamental rights impact assessments, transparency requirements, and the duty to ensure meaningful human oversight. These mechanisms are to be welcomed as a first step toward aligning technological innovation with the principles of administrative law.

Nevertheless, the analysis has also identified several critical limitations in the current regulatory framework. The AI Act's emphasis on system-level risk classification may fail to capture the contextual dimensions of administrative decision-making. Its concept of meaningful human oversight remains under-defined and may prove inadequate in the face of automation bias, limited institutional capacity, and the frenzy of digitalisation (Pavlidis, 2021). Although the AI Act introduces transparency obligations, it does not provide individuals with a fully-fledged right to explanation, nor does it establish dedicated procedures for challenging algorithmic decisions. Much will depend on how the provisions of the AI Act are implemented and complemented within national administrative and judicial systems.

In light of these findings, a number of recommendations can be proposed. First, the concept of meaningful human oversight must be clarified and operationalised through detailed guidance and sector-specific implementation practices, ensuring intervention and accountability. Second, procedural safeguards should be strengthened through national legislation that provides clear and enforceable rights to receive an explanation and to access effective remedies in cases of automated administrative decision-making. Third, fundamental rights impact assessments must be given greater practical relevance by requiring stakeholder participation, external review, and public disclosure of results. Fourth, the principle of proportionality should inform both the design and the enforcement of AI-related obligations in the public sector, ensuring that compliance burdens are commensurate with both the abstract risk posed by a system and the practical realities of its deployment. Finally, public authorities and supervisory institutions must invest in building the technical, legal, and institutional capacity required to govern AI in a manner that is rights-respecting.

The future of algorithmic governance in the public sector will ultimately depend on the development of dynamic oversight frameworks capable of responding to technological change without compromising democratic principles (Wirtz, Weyerer, and Sturm, 2020). This requires not only the periodic review and refinement of the AI Act, but also the cultivation of a jurisprudence on algorithmic legality, grounded in administrative law and fundamental rights. In the end, the challenge is not only to regulate AI, but to ensure that public power remains subject to law, when exercised through or with the assistance of algorithms.